\documentclass[a4paper, twocolumn,superscriptaddress]{revtex4-1}

\usepackage{amsmath,amssymb,amsfonts} 
\usepackage{bm} %bold math and greek
\usepackage{graphicx}
\usepackage{textcomp} 
\usepackage{dsfont}

\newcommand{\LOA}{Laboratoire d'Optique Appliqueé, CNRS, Ecole Polytechnique, ENSTA Paris, Institut Polytechnique de Paris, 181 chemin de la Hunière et des Joncherettes 91120 Palaiseau, France}
\newcommand{\LLG}{Laser-Laboratorium Göttingen e.V., Hans-Adolf-Krebs-Weg 1, 37077 Göttingen, Germany}
\newcommand{\MBI}{Max Born Institute for Nonlinear Optics and Short Pulse Spectroscopy, Max-Born-Strasse 2A, 12489 Berlin, Germany}

\begin{document}

\title{Relativistic near-single-cycle optics at 1 kHz}

\author{Marie Ouill\'e}\affiliation{\LOA}
\author{Aline Vernier}\affiliation{\LOA}
\author{Frederik Boehle}\affiliation{\LOA}
\author{Maimouna Bocoum}\affiliation{\LOA}
\author{Magali Lozano}\affiliation{\LOA}
\author{Jean-Philippe Rousseau}\affiliation{\LOA}
\author{Zhao Cheng}\affiliation{\LOA}
\author{Domynikas Gustas}\affiliation{\LOA}
\author{Andreas Blumenstein}\affiliation{\LLG}
\author{Peter Simon}\affiliation{\LLG}
\author{Stefan Haessler}\email{stefan.haessler@ensta-paris.fr}\affiliation{\LOA}
\author{J\'er\^ome Faure}\email{jerome.faure@ensta-paris.fr}\affiliation{\LOA}
\author{Tamas Nagy}\email{nagy@mbi-berlin.de}\affiliation{\MBI}
\author{Rodrigo Lopez-Martens}\email{ rodrigo.lopez-martens@ensta-paris.fr}\affiliation{\LOA}

\begin{abstract}
We present a laser source delivering waveform-controlled 1.5-cycle pulses that can be focused to relativistic intensity at 1\,kHz repetition rate. These pulses are generated by nonlinear compression of high-temporal-contrast sub-25\,fs pulses from a kHz Ti:Sapphire double-chirped pulse amplifier in a stretched flexible hollow fiber compressor scaled for high peak power. The unique capabilities of this source are demonstrated by observing carrier-envelope phase effects in laser-wakefield acceleration of relativistic electrons for the first time.

\end{abstract}

\maketitle
%%%%%%%%%%%%%%%%%%%%%%%%%%%%%%%%%%%%%%%%%%%%%%%%%%%%%%%%%%%%%%%%%%%%%%%%%%%%%%%%%%%%

\section{Introduction}
\label{sec:intro}

\begin{figure*}
	\centering
		\includegraphics[width=\linewidth]{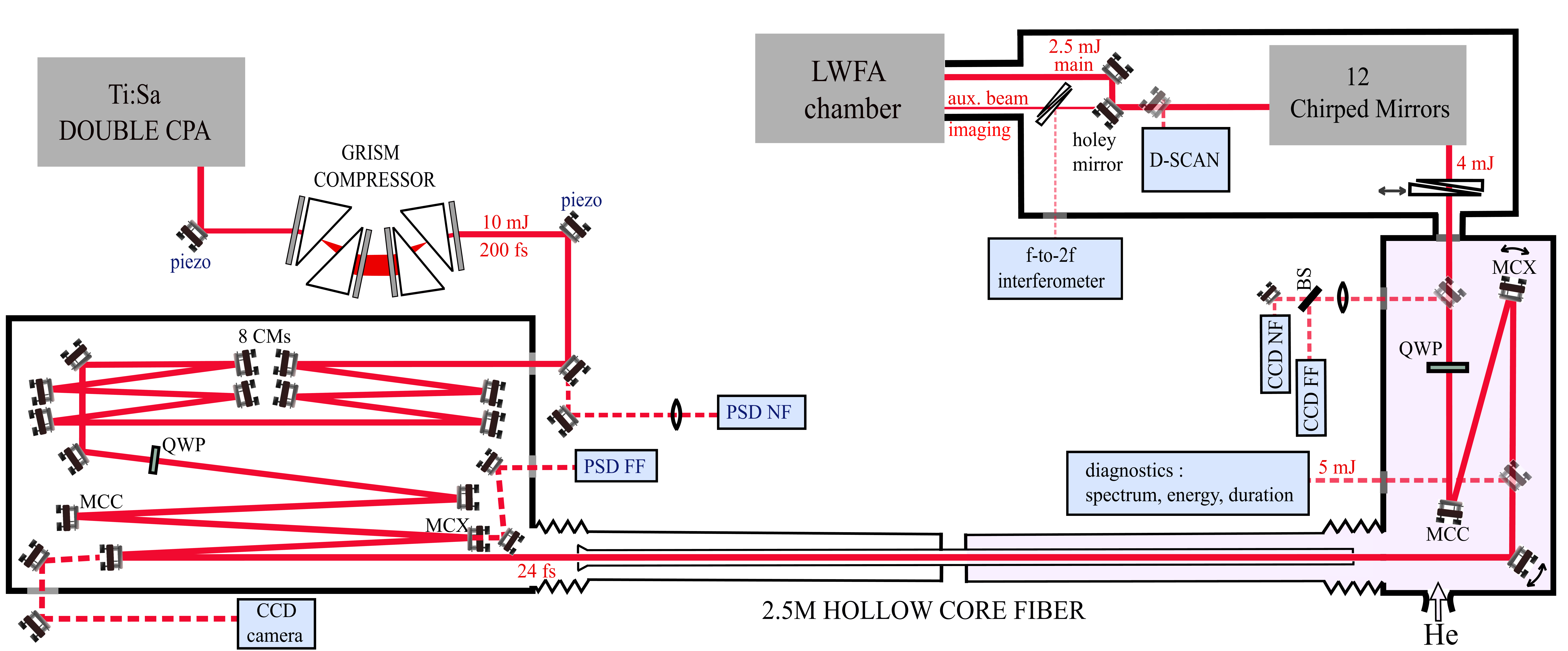} %[width=0.9\textwidth]
	\caption{Schematic of the vacuum-integrated stretched flexible hollow fiber compressor pulse compressor setup. "PSD": photosensitive detector; "NF": Near Field; "FF": Far Field; "piezo": piezo driven mirror mounts; "MCX": convex mirror; "MCC": concave mirror; "QWP": Quarter Wave Plate}
	\label{fig:Figure1}
\end{figure*}
 
Based on Ti:sapphire femtosecond chirped pulse amplifier (CPA) technology and hollow-core fiber (HCF) nonlinear post-compression~\cite{nisoli_generation_1996}, laser transients approaching a single carrier-wave period with controlled carrier-envelope phase (CEP) have been available for more than a decade with sufficient average ($<$ W) and peak power ($\leq$ 0.1\,TW) to drive laser-atom interactions in the strong-field regime with unprecedented temporal precision, thereby paving the way for attosecond science \cite{krausz2,calegari2016,ciappina2017}.  

Recent developments have led to significant increase in the generated power of sub-2-cycle laser pulse sources. On the one hand, Ytterbium-based fiber CPA technology combined with HCF post-compression has enabled the scaling of the average power to >200~W~\cite{hadrich2016} and the increase in achievable compression factors (33-fold)~\cite{jeong2018}. For such systems, however, CEP-locking has yet to be implemented. On the other hand, broadband optical parametric chirped pulse amplifiers (OPCPA) have succeeded in scaling up the peak power of few-cycle pulses to the multi-TW range at the cost of CEP-locking due to a low (10-Hz) repetition rate~\cite{gruson_fopa_2017,rivas2017,kessel2018}. While progress in high repetition rate pump-laser technology and passively CEP-stable seed generation~\cite{baltuska_controlling_2002} for OPCPA now allows for higher average power and good CEP stability~\cite{budriunas2017}, this has yet to be united with sub-2-cycle duration. 

A major motivation for increasing the peak power of few-cycle lasers is the study of relativistic laser-plasma interactions, where the  field-driven electron quiver energy exceeds its rest mass energy ($\approx 0.5$~MeV). Because of their inherently high temporal contrast, OPCPA-based sources have recently enabled such experiments at 10-Hz repetition rate, demonstrating relativistic surface high-harmonic generation with sub-2-cycle~\cite{rivas2017} and sub-3-cycle~\cite{kormin2018,kessel2018,jahn2019} pulses, as well as laser-wakefield acceleration (LWFA) with 3-cycle pulses~\cite{schmid2009}.  

When it comes to further reducing the driving pulse duration towards the single-cycle limit at kHz repetition rate, Ti:sapphire CPA systems paired with HCF post-compression remain serious competitors~\cite{Jarque_universal_route_2018} and relativistic LWFA has recently been demonstrated with such a driver~\cite{guenot_relativistic_2017}. TW peak powers have been demonstrated for 5-fs pulses~\cite{bohman2010} from such a setup. The first system uniting multi-mJ pulse energy, sub-2-cycle duration and CEP-locking~\cite{boehle_compression_2014} was the precursor to the system described here. 

Our advanced setup is the result of a large engineering effort resulting in significant performance improvements: the pulses set a new duration record (1.5 cycle) for TW~peak~power pulses and the pulse stability (CEP, energy, spatial, spectral) fulfills the requirements for relativistic-intensity laser-matter experiments. Together with the kHz repetition rate, high temporal contrast ratio and achievable ultra-high intensity on target, this system is now uniquely suited for driving relativistic-intensity light-matter interactions with sub-cycle time control over the driving light waveform. Here, we report on the first experimental observation of CEP effects in relativistic LWFA driven in the near-single-cycle regime.

\section{Power-scaled hollow fiber compressor}
\label{sec:HCF}

The fully vacuum-integrated post-compressor setup is shown in detail in Fig.~\ref{fig:Figure1}. The seed laser consists of a kHz Ti:sapphire double CPA chain delivering 24-fs pulses with 10\,mJ energy (shot-to-shot stability of < 0.3~\%rms over hours) at a 1-kHz repetition rate with a temporal contrast of $>10^{10}$ at $\approx-10$\,ps before the pulse peak~\cite{jullien_SN2doubleCPA_2014}. Pulses from the CPA are partially compressed down to $\approx 200\:$fs in air so as to prevent nonlinear beam degradation in the entrance window of the vacuum beamline, which would significantly reduce the coupling efficiency into the HCF~\cite{boehle_compression_2014}. Final compression down to the Fourier-transform limited duration of 24\,fs is achieved after 8 highly dispersive chirped mirrors under vacuum introducing $\approx -2000\:$fs$^2$ group delay dispersion (GDD).

The post-compression stage is based on self-phase-modulation (SPM) in a gas-filled HCF, a now widely used technique~\cite{nisoli_generation_1996} that has the advantage of producing well-compressible broadband pulses with excellent beam profile and spatially homogeneous spectrum~\cite{agrawal_book}. We implemented a combination of three strategies to scale this technique to the $\approx$ 0.4\,TW peak power of our CPA chain. 

First, we exploit stretched flexible hollow-core fiber technology enabling arbitrary waveguide length without degradation of the waveguiding properties~\cite{nagy_flexible_2008,nagy_optimal_2011}. The HCF dimensions were scaled to 2.5\,m length and 536\,$\mu$m inner diameter. A conical glass taper is coaxially installed at the HCF entrance to create a very robust protection against damage due to slight misalignments or pedestals in the spatial beam profile. The beam is focused into the fiber using a reflective mirror telescope with an effective focal length of $\approx$ 4.2\,m and optimal coupling into the fiber is maintained with active beam pointing stabilization to ensure long-term stability. Second, high-purity helium gas is differentially pumped through the HCF, thus forming a stable pressure gradient~\cite{suda_generation_2005} across the waveguide. The pressure gradually increases from $<1\:$mbar at the fiber entrance to the static filling pressure of the output chamber up to 2\,bar. This prevents undesirable nonlinear phenomena around the fiber entrance and enhances the coupling efficiency and stability. Furthermore, the increasing pressure counteracts the decreasing nonlinearity due to propagation losses inside the fiber. 
Third, both multi-photon ionization and self-focusing are further mitigated by using circular polarization~\cite{ghimire_high-energy_2005,chen_generation_2009}, which also reduces losses and instabilities due to cycling of energy between fiber modes~\cite{malvache_multi-mj_2011}. 
Two broadband quarter-wave plates are therefore placed before and after the HCF to switch the laser polarization between linear and circular. 

\begin{figure}[hbt!]
	\centering
		\includegraphics[width=\linewidth]{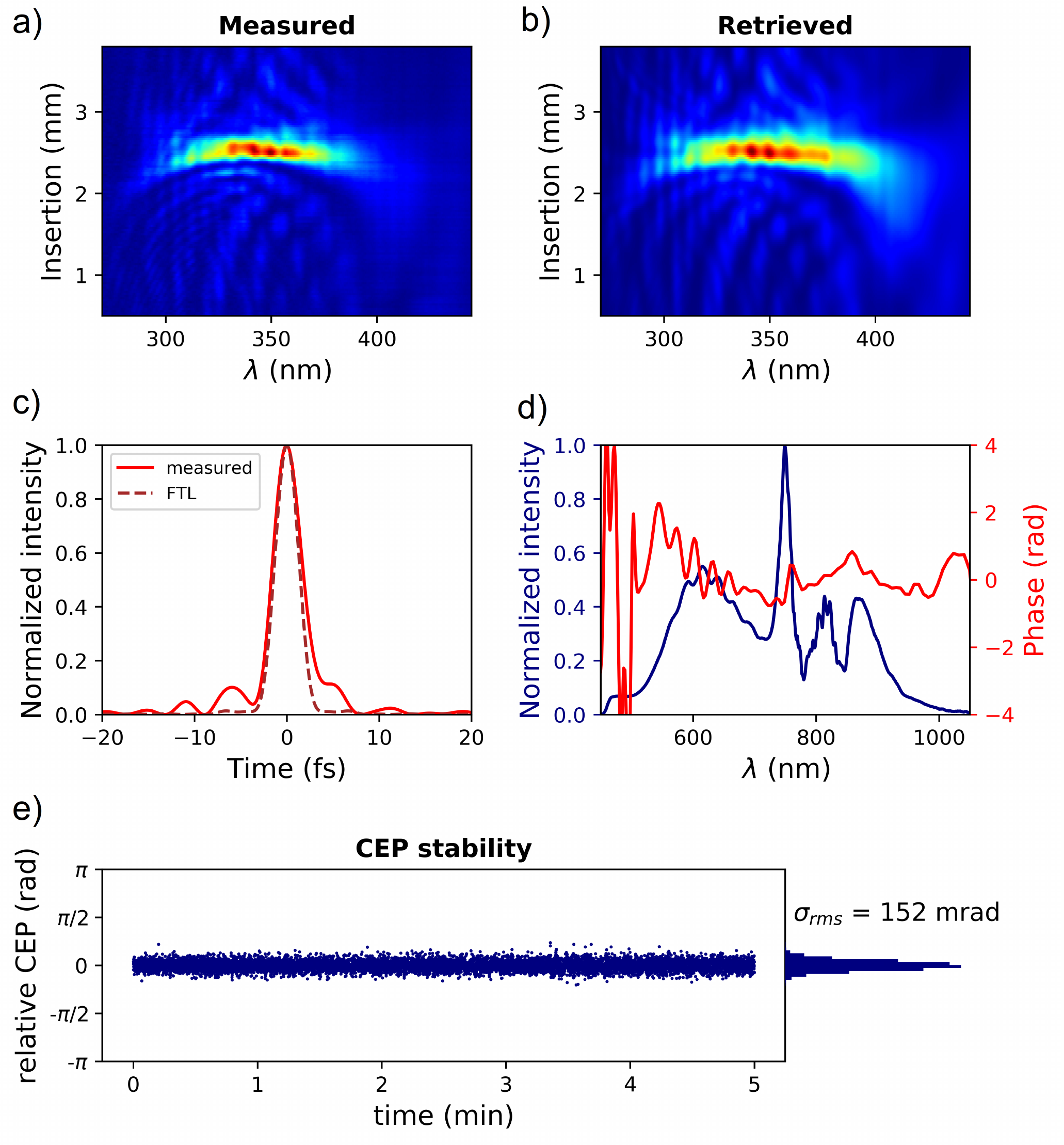}
		\caption{D-scan measurement for $1.3\:$bar helium pressure. Measured (a) and retrieved (b) d-scan traces. Reconstructed temporal profile (c): the retrieved temporal pulse profile is shown in red (3.6\,fs FWHM) and the ideally compressed pulse (flat phase, 2.9\,fs FWHM) is shown by the brown dotted line. Measured spectral intensity and phase (d). Relative CEP stability measured over 5 min (each point is averaged over 30 shots) (e).}
	\label{fig:Figure2}
\end{figure}

In the output chamber, two insertable mirrors can send the beam to diagnostics for position, spatial profile, spectrum and pulse energy. Two active mirrors then allow for alignment onto the near- and far-field references thus obtained. A 3-mm thick fused-silica window separates the helium-filled output chamber from the vacuum-beamline downstream. The beam first propagates through a pair of motorized fused silica wedges. The positive group delay dispersion (GDD) induced by SPM in the fiber as well as the propagation through the quarter-wave plate and the window is slightly over-compensated for with the six pairs of highly dispersive ($-40\:\mathrm{fs}^2$ each) double-angle chirped mirrors and the wedge position is tuned for fine adjustment of the GDD leading to optimum pulse compression. A dispersion-scan device~\cite{miranda_characterization_2012} (Sphere Ultrafast Photonics) placed under vacuum serves as a precise temporal measurement immediately before the experimental chamber with controllable dispersion provided by precise insertion of one of the wedges. With $1.3\:$bar helium pressure in the output chamber, the minimum achievable FWHM pulse duration with our system is 3.6 $\pm$ 0.1\,fs (see Fig. \ref{fig:Figure2}c). This duration is close to the Fourier limit of 2.9\,fs (with 60\,\% higher peak power) supported by the broadened spectrum.
This is a remarkable result indicating accurate compensation of not only the GDD but also the third-order dispersion (TOD), to which compression in the sub-two optical cycle regime is extremely sensitive~\cite{schmidt_compression_2010,suda_effects_2012,Jarque_universal_route_2018}. As a consequence, venting the d-scan chamber with air is sufficient to spoil the compression shown in Fig.~\ref{fig:Figure2} and the optimal wedge position leaves a significant negative TOD on the compressed pulses, similar to the result reported in Ref.~\cite{timmers_generating_2017}. This is due to the different GDD/TOD ratio of air compared to that of the fused silica wedges.

The remaining room for improvement on the result in Fig.~\ref{fig:Figure2} is left by two imperfections in the compression. The first and minor one is a very small remaining third-order-dispersion of $\approx 5\:$fs$^3$, without which the pulse would be 3.3~fs long with 10\% higher peak power. The second and dominating one is the phase distortion introduced by the chirped mirrors on the blue end of the spectrum. The phase oscillations between 700 and 550~nm reduce the energy contained in the main peak and smoothing them (possible by using complementary-pair CMs instead of double-angle CMs) would increase the peak power by 15\% without changing the pulse duration. In order to further approach the Fourier limit, the uncontrolled phase below 520~nm would need to be smoothed and flattened, which would require a more advanced and to our knowledge not yet available CM design. This would decrease the pulse duration to 3~fs and increase the peak power by 25\%. 

Energy measurements performed for 1.3\,bar of helium with a single-shot energy detector (noise level $\approx$ 10\,$\mu$J) yield an excellent pulse-to-pulse stability of $\approx 0.4\,\%$ rms and typical pulse energies of 4.5$-$5\,mJ right after the fiber, 3.5\,mJ after the chirped-mirror compressor at the entrance of the d-scan device and 2.5\,mJ on target in the LWFA chamber. Losses along the beamline are due to the wave plate, the wedges ($\approx 5\,\%$), the CMs ($\approx 5\,\%$ from all 12 CMs) and the transport mirrors ($\approx 1.5\,\%$ per mirror).

CEP stabilization of the system is based on a home-made f-to-2f interferometer. As shown in Fig.~\ref{fig:Figure1}, the f-to-2f signal is generated using the reflection from the front face of a thin wedge pair placed in the beam path of an auxiliary beam passes created with a holed mirror and used for plasma imaging~\cite{guenot_relativistic_2017}. The derived error signal then modulates the phase offset of the oscillator locking electronics to correct for slow CEP drifts accumulated through the laser chain. As shown in Fig. \ref{fig:Figure2}e, the residual CEP noise of the system over 5 minutes is $\approx 150\:$mrad rms.

\section{Pulse duration tuneability}
\label{sec:tune}

\begin{figure}
	\centering
		\includegraphics[width=\linewidth]{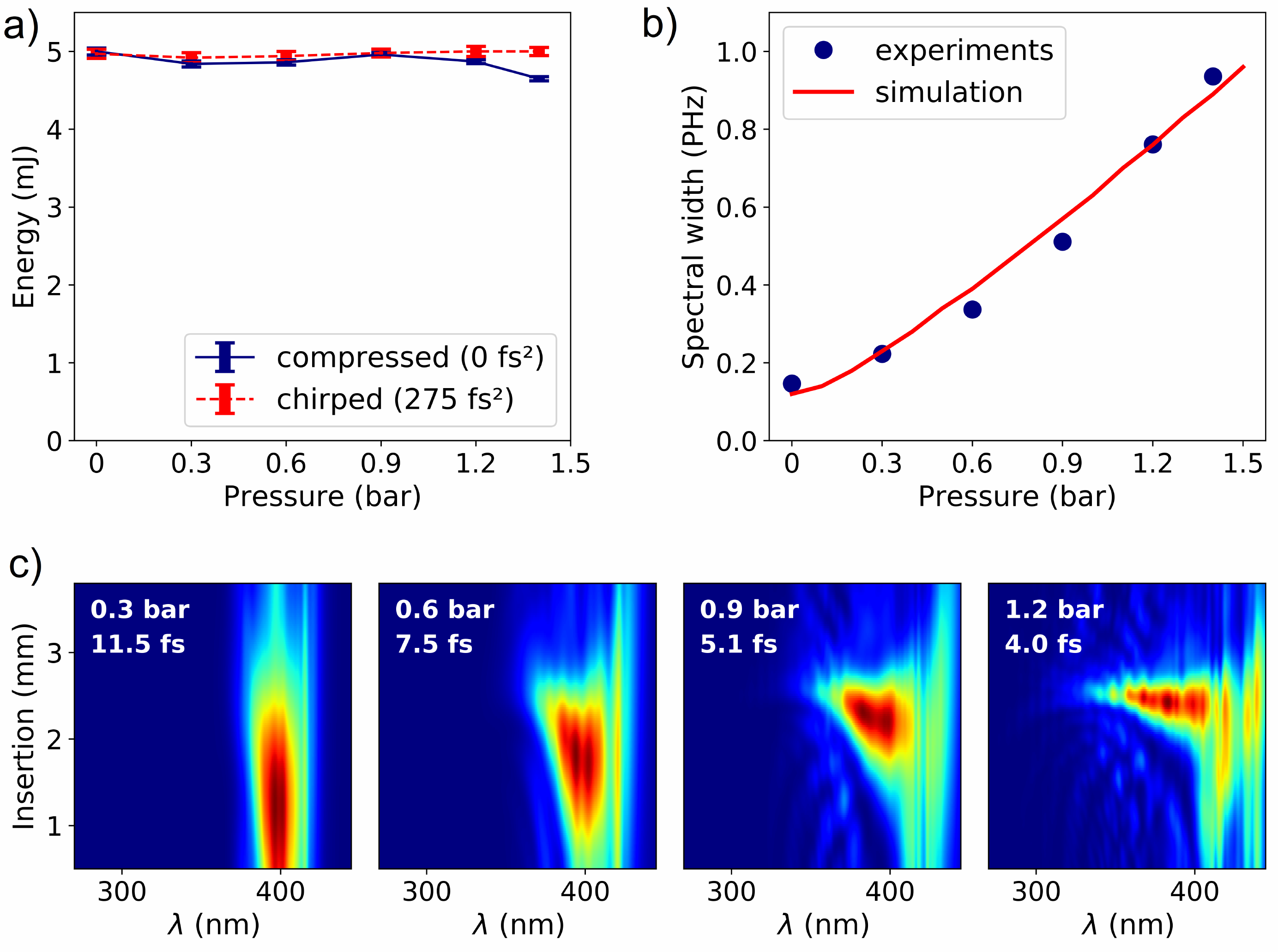}  
		\caption{(a) Evolution of the output energy with the gas pressure for a compressed pulse (blue) and a positively chirped pulse (red) showing that we do not have ionization-induced losses, except at the highest pressures. (b) The spectral width evolution while tuning the helium pressure in the output chamber (blue dots) is in good agreement with numerical simulations taking only Kerr nonlinearities into account (red line). (c) D-scan traces for different pressure values showing the pulse duration tunability of the laser.}
	\label{fig:Figure3}
\end{figure}

Compression data for different gas pressures (Figure \ref{fig:Figure3}) show that our HCF post-compression stage is adequately energy-scaled, i.e.~self-focusing and ionization-induced effects are avoided: the fiber transmission is the same for a Fourier-limited input pulse as for a positively chirped ($+275\:$fs$^2$) input pulse; the fiber transmission remains the same over the complete helium-pressure range used. Only at the highest pressure of 1.4~bar, which is above our usual working range, does the transmission start to drop. Finally, as shown in Fig.~\ref{fig:Figure3}b, the measured spectral broadening follows the helium pressure $p$ as expected for purely SPM-induced broadening. We compare the experimental spectral widths (defined as twice the rms-bandwidth $\sigma_\omega = \sqrt{\langle\omega^2\rangle - \langle\omega\rangle^2}$ ) to values obtained from a numerical solution of the one-dimensional generalized nonlinear Schrödinger equation ~\cite{brabec_nonlinear_1997,deiterding_reliable_2013} for a Kerr nonlinearity. These simulations start from an experimentally measured input pulse, obtained from a Wizzler measurement (Fastlite~\cite{Oks2010}) located after the diagnostics port of the output chamber (cf. Fig.~\ref{fig:Figure1}) for an evacuated HCF, and describe dispersion, SPM and self-steepening, the latter of which significantly modulates SPM-induced spectral broadening for input pulses as short as ours. The pressure gradient along the HCF of length $L=2.5\:$m was modeled as $p(z) = p_\mathrm{L} \sqrt{z/L}$~\cite{suda_generation_2005}, where $p_\mathrm{L}$ is the pressure at the HCF exit located at $z=L$. Neither ionization-induced effects nor spatial effects had to be included in these 1D-simulations in order to obtain satisfactory agreement with the experimental spectra. Note that the functional shape of the pressure dependence remains very close to that derived for pure SPM and Fourier-transform-limited Gaussian input pulses~\cite{pinault_frequency_1985}.

An interesting consequence of this well-scaled post-compression stage is that the pulse duration can be easily tuned from 25~fs down to 3.5~fs by simply varying the gas pressure and adjusting dispersion, while keeping a constant output pulse energy (see Figs.~\ref{fig:Figure3}c and~\ref{fig:Figure2}) and similar spatial beam profile imposed by the waveguide.

\section{CEP-dependence of relativistic laser-wakefield acceleration}
\label{sec:examples}

\begin{figure}[hbt!]
\centering
		\includegraphics[width=\linewidth]{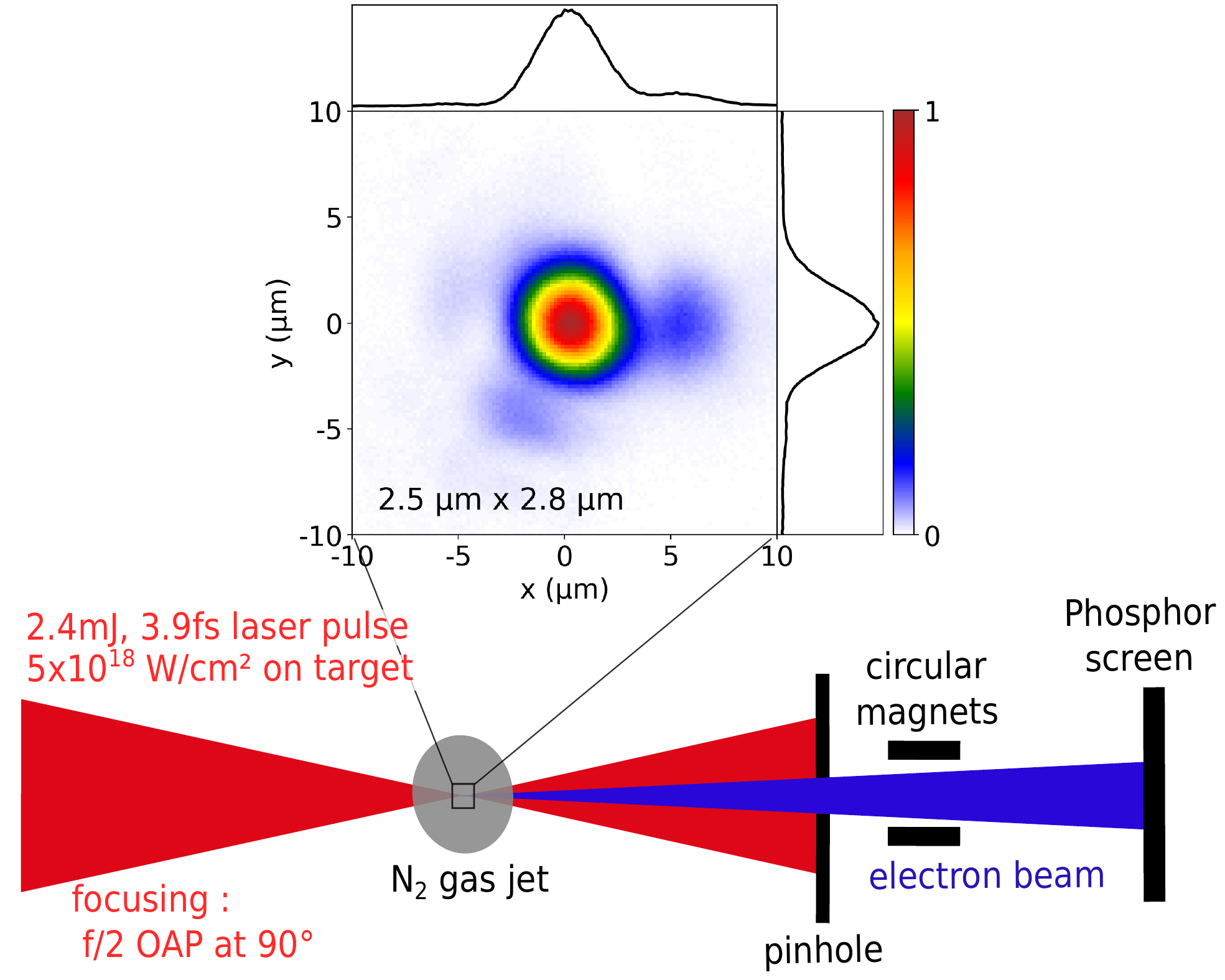}  
		\caption{Experimental setup for electron acceleration (top view) and picture of the laser focal spot with size 2.5 x 2.8\,$\mu$m FWHM, resulting in intensities of $\approx 5\times10^{18}$ $\mathrm{W \cdot cm^{-2}}$. }
	\label{fig:Figure4}
\end{figure}

As a demonstration of the excellent spatiotemporal quality of our system, we now focus on their application to LWFA in the relativistic-intensity regime. The experimental setup for LWFA is shown in Fig.~\ref{fig:Figure4}. 
Laser wakefield acceleration of electrons driven by near-single cycle pulses was demonstrated with a similar setup but without CEP stability~\cite{guenot_relativistic_2017,Gustas2018}. The beam is focused into a continuously flowing microscale supersonic nitrogen gas jet using a  90\,\textdegree off-axis f/2 parabola. We obtain a near-Gaussian 2.5 $\times$ 2.8~µm (FWHM) focal spot (see Fig. \ref{fig:Figure4}), corresponding to an on-target intensity of $\approx 5\times 10^{18}$~$\mathrm{W \cdot cm^{-2}}$ with 2.4~mJ on target (the relativistic limit for which the normalized vector potential reaches $a_0=1$ is $\approx 2.6 \times 10^{18}$ $\mathrm{W \cdot cm^{-2}}$ at 719\,nm central wavelength). More details on the electron detection system can be found in ref.~\cite{Gustas2018}

\begin{figure}[hbt!]
\centering
		\includegraphics[width=\linewidth]{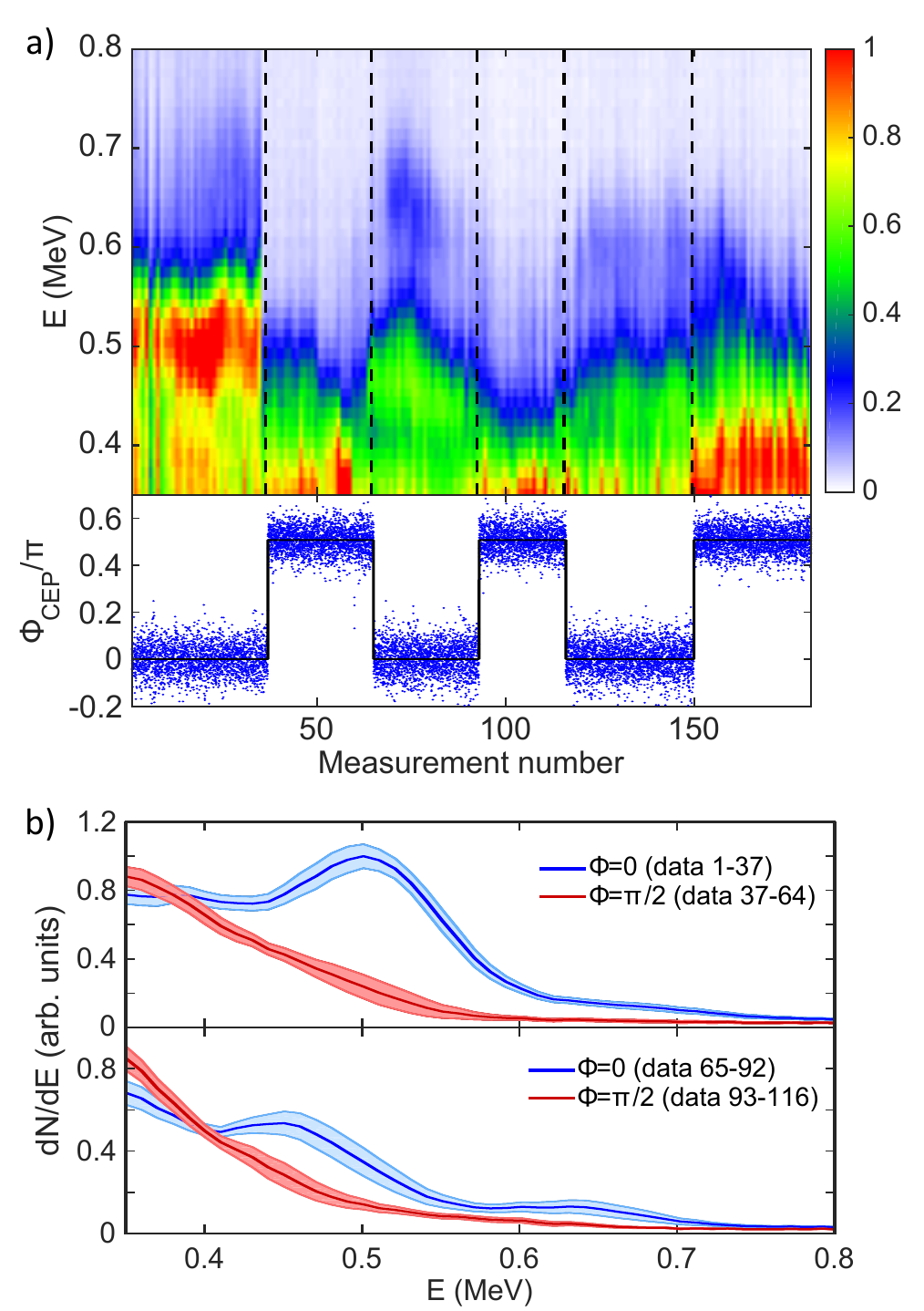} 
\caption{a) Top: Electron spectra $\mathrm{d}N/\mathrm{d}E$ (arb. u.) obtained while varying the CEP of the accelerating laser pulse (each spectrum was obtained by averaging over 500 laser shots). Bottom: the measured CEP (blue dots) and the command CEP values sent to the feed-back loop (black line). b) Averaged spectra for the first (top) and second (bottom) CEP cycle. The solid dark lines represent the average spectrum whereas the light blue and red areas indicate the standard deviation due to spectral fluctuations.}        
	\label{fig:Figure5}
\end{figure}

In these experiments, the laser pulse envelope drives the wakefield via the ponderomotive force. Electrons are injected into the wakefield in a process known as ionization injection~\cite{pak10,mcgu10}. In ionization injection, electrons are born at the peak of the laser electric field and subsequently injected and accelerated into the wakefield accelerating structure. This injection process heavily depends on the CEP because the laser phase determines the initial conditions of electrons in the longitudinal phase space, which eventually impacts their final momenta~\cite{lifschitz2012}. More precisely, the CEP controls (i) the amplitude of the most intense laser half-cycle and therefore the number of injected electrons and (ii) the exact initial conditions of the trapped electrons, which impact the final electron energies and angles. Observing such CEP effects in the experiment is rather challenging because the CEP varies rapidly during laser propagation since the laser phase velocity ($v_{\varphi}$) and group velocity ($v_g$) are different in the plasma. Typically, the CEP spans $2\pi$ over the phase splippage length $L_{2\pi}=\lambda c/(v_\varphi-v_g)\approx \lambda n_c/n_e$, where $n_e$ and $n_c$ are the electron plasma density and critical density. For our experimental parameters, $n_e/n_c\approx 0.1$ and $L_{2\pi}\approx 8$~µm. Therefore, the effect of CEP on electron injection is significant only if the injection length is smaller than $L_{2\pi}$, which therefore requires a very localized injection and places stringent demands on the stability of all other pulse parameters in space, time and energy, met by our system for the first time. 

Despite these difficulties, a clear CEP signature could be observed on the electron energy distribution. Figure~\ref{fig:Figure5}a) shows a cascade plot of successive electron spectra recorded as the CEP was cycled from 0 to $\pi /2$. Note that the absolute value of the CEP is not measured, only the relative changes are known. For a relative CEP of 0, the spectrum exhibits a peak close to 0.5\,MeV and the distribution tail displays another smaller feature at 0.65\,MeV. These features disappear when the CEP is shifted by $\pi/2$. While the effect of CEP is initially clear (data 1 to 125), it should be noted that the difference tends to wash out towards the end of the scan, indicating that other parameters might be changing as well, such as the gas density. However, the effect of CEP is clearly demonstrated by looking at the averaged spectra and comparing the results for 0 and $\pi/2$ relative CEP values, see Fig.~\ref{fig:Figure5}b). The CEP effect clearly outweighs that of the spectral fluctuations as it is significantly larger than the standard deviation of the spectra. These results constitute the first observation of CEP effects in underdense relativistic laser-plasma interactions and shall be investigated further.

To conclude, our laser system delivers pulses combining high peak powers up to 1\,TW, near-single-cycle pulse duration (3.6\,fs or 1.5 cycle at 719\,nm central wavelength), excellent beam quality and high temporal contrast inherited directly from the double CPA seeding the HCF post-compressor stage. Furthermore, the output pulses exhibit the required spectral, energy and CEP stability for the investigation of relativistic-intensity laser-matter interactions on sub-light-cycle time scale. We observe for the first time carrier-envelope phase effects in laser wakefield acceleration demonstrating the unique capabilities of our light source.

\section*{Funding}
Agence Nationale pour la Recherche (ANR-11-EQPX-005-ATTOLAB, ANR-14-CE32-0011-03 APERO); Laboratoire d'Excellence Physique: Atomes Lumi\`ere Mati\`ere (LabEx PALM) overseen by the Agence Nationale pour la Recherche as part of the \emph{Investissements d'Avenir} program (ANR-10-LABX-0039); European Research Council (ERC Starting Grant FEMTOELEC 306708), LASERLAB-EUROPE (grant agreement no. 284464), R\'egion Ile-de-France (SESAME 2012-ATTOLITE).

\bibliography{SN2laser_forarXiv}

\end{document}